\documentclass{ws-mpla}

\usepackage{bm}
\usepackage{amsmath}
\usepackage{amssymb}

\newcommand*{\cm}{center of mass }
\newcommand*{\brho}{{\bm{\rho}}}
\newcommand*{\hrho}{{\hat{\bm{\rho}}}}
\newcommand*{\ie}{i.e., }
\newcommand*{\wfn}{wave function }
\newcommand*{\ba}{{\mathbf a}}

\newcommand*{\bk}{{\mathbf k}}
\newcommand*{\br}{{\mathbf r}}

\newcommand*{\bx}{{\mathbf x}}

\newcommand*{\bq}{{\mathbf q}}

\newcommand*{\hk}{{\hat{\mathbf k}}}

\newcommand*{\teff}{{\text{eff}}}

\newcommand*{\tas}{{\text{as}}}
\newcommand*{\ti}{{\text{i}}}
\newcommand*{\tf}{{\text{f}}}
\newcommand*{\tif}{{\text{if}}}
\newcommand*{\tC}{{\text{C}}}
\newcommand*{\tN}{{\text{N}}}

\newcommand*{\texp}{{\text{exp}}}
\newcommand*{\tprior}{{(\text{prior})}}
\newcommand*{\tpost}{{(\text{post})}}
\newcommand*{\tDWBA}{{\text{DWBA}}}
\newcommand*{\tBe}{{\textrm{Be}}}
\newcommand*{\tB}{{\textrm{B}}}
\newcommand*{\tPb}{{\textrm{Pb}}}
\newcommand*{\tLi}{{\textrm{Li}}}
\newcommand*{\tBep}{{}^7\textrm{Be}p}
\newcommand*{\be}{\begin{equation}}
\newcommand*{\ee}{\end{equation}}
\newcommand*{\bay}{\begin{eqnarray}}
\newcommand*{\eay}{\end{eqnarray}}
\newcommand*{\B}{${}^8\textrm{B}$ }

\newcommand*{\Pbr}{${}^{208}\tPb({}^{8}\tB,{}^{7}\tBe\, p){}^{208}\tPb$ }
\newcommand*{\Be}{${}^7\textrm{Be}$ }
\newcommand*{\Bep}{${}^7\textrm{Be-}p$ }
\newcommand*{\SE}{Schr\"odinger equation }

\begin{document}

\title{THREE-BODY COULOMB FINAL-STATE INTERACTION EFFECTS IN THE COULOMB
BREAKUP OF LIGHT NUCLEI}

\author{\footnotesize E. O. ALT}
\address{\it Institut f\"ur Physik, Universit\"at
Mainz, D-55099 Mainz, Germany\\
Erwin.Alt@uni-mainz.de}

\author{\footnotesize B. F. IRGAZIEV}
\address{\it Theoretical Physics
Department, National University, Tashkent 700174, Uzbekistan\\
qcd@uzsci.net}

\author{\footnotesize A. M. MUKHAMEDZHANOV}
\address{\it Cyclotron Institute, Texas A\&M
University, College Station, TX 77843, USA \\
akram@comp.tamu.edu}

\maketitle

\begin{center}
Preprint MZ-TH/05-06
\end{center}

\begin{abstract}
Coulomb breakup of a projectile in the Coulomb field of a fully
stripped heavy nucleus is at present one of the most popular
experimental methods to obtain information on reactions of interest
in nuclear astrophysics. Its theoretical interpretation presents,
however, considerable difficulties, due to the three-body nature and
the infinite range of the Coulomb forces involved. Among the
uncertainties affecting present analyses, the possible modification
of the dissociation cross section by three-body Coulomb final-state
interactions plays a major role. Various methods which have been
proposed to deal with it are briefly reviewed. However, none of them
is based on a consistent and mathematically satisfactory quantum
mechanical treatment, with the exception of the one proposed
recently. The latter, being based on the prior form of the
dissociation amplitude, makes use of a genuine three-charged
particle wave function for the final state which is an exact
solution of the three-body Schr\"odinger equation, albeit only
asymptotically, \ie for large distances. Nevertheless, interesting
conclusion can be drawn concerning the influence of three-body
Coulomb final-state interactions on quantities of astrophysical
interest.

\end{abstract}

\ccode{PACS Nos: 25.70.Hi, 25.70.Bc, 24.10.Eq, 24.10.Ht}


\section{Introduction}
\protect\label{introd}

Coulomb breakup reactions, being three- (or more-) problems, are
dynamically very complicated due to the influence of strong
interactions acting between the colliding nuclei; and the long-range
character of the Coulomb interaction adds non-trivially to the
complication. However, in the special case of very peripheral
collisions the nuclei do not come close to each other so that the
short-range nuclear forces between them can be neglected. Then the
reaction mechanism simplifies considerably: the nuclei interact with
each other through the electromagnetic field only and unambiguous
conclusions can be drawn from such experiments. Consider the
dissociation
\begin{equation}
a + A \to b + c + A, \label{brkp1}
\end{equation}
where particle $a$ is the bound state $a=(bc)$, and $A$ is a heavy
target. The basic mechanism underlying Coulomb breakup is as
follows: the fast projectile $a$ moves in the Coulomb field of the
target $A$, with the impact parameter being larger that the sum of
the radii of the nuclei $a$ and $A$. The electromagnetic target
field then lifts the fragments $b$ and $c$ into the continuum. There
they interact with each other via the sum of the Coulomb plus
nuclear potentials, but with the target only Coulombically.

Thus, by choosing appropriate kinematic conditions Coulomb breakup
represents a powerful, though indirect, tool to obtain information
on the radiative capture cross section $b+c \to a+\gamma$, at
astrophysically relevant energies\cite{baur86} (we also recommend
the reader to consult Ref.~\cite{baur03} where the general theory of
Coulomb excitation and dissociation and their applications are
reviewed). Its advantages are exploited around the world in
different nuclear centers (GSI, KfK, Germany,\cite{CDR,kiener91}
GANIL, France,\cite{lef95} NSCL, MSU, Cyclotron Institute, Texas
A\&M University, USA,\cite{riawhitepaper,riaweb,austin2001,uts90}
RIKEN, Japan.\cite{ribfweb}) It is, however, clearly impossible to
cover in a short review all experiments where Coulomb breakup
reactions have been used to obtain astrophysical information.

Theoretically, Coulomb breakup reactions are studied, e.g., by post
and prior forms of the Distorted-Wave Born Approximation
(DWBA),\cite{baur84,rebel90,bertulani91} the semiclassical
approach,\cite{baur94} the diffractive dissociation
approach,\cite{heide89} and the continuum-discretized coupled
channel (CDCC) method,\cite{yahiro84,mortimer02,matsumoto03} each of
which has specific advantages and disadvantages. It must, however,
be kept in mind that the astrophysical factors (``$S$-factors'')
extracted from the analysis of Coulomb breakup data are affected by
various uncertainties due to (i) the unavoidable extrapolation of
the data to astrophysically effective energies, (ii) contributions
from various electromagnetic multipoles ($E1, E2, M1$), (iii)
assumptions about the nuclear interactions, and (iv) various
higher-order effects (see Refs.
\cite{Norbury93,esbensen95,bertulani95,bertulani96,typel01} and
references therein).

One of the most interesting higher-order effects is the so-called
post-decay acceleration (PDA) of the fragments in the Coulomb field
of the target. It results from the fact that, if the fragments of
the projectile have different charge-over-mass ratios, the repulsive
force exerted by the target nucleus will give rise to a different
acceleration of each fragment (PDA) which eventually leads to a
change in their relative energy on their way from the breakup point
to the detector. Consequently, the experimentally accessible
amplitude of the projectile photodissociation will differ from the
physically interesting one. One of the first direct
measurements\cite{uts90} of the PDA has been performed for the
Coulomb breakup $^7\textrm{Li}\to \alpha +t$ on ${}^{208}$Pb, at a
projectile incident energy $E_{\ti}=63$ MeV. There it has been been
found that the minimum of the coincidence spectrum of
$\alpha$-particles and tritons which is expected to occur at zero
relative energy $E_{\alpha t}$, was shifted due to the different
acceleration of each fragment in the Coulomb field of the target.
Evidently, this different PDA of the $\alpha$ and the triton is due
to their different charge-over-mass ratios. The shift of the minimum
in the coincidence spectrum can serve as an nuclear
clock.\cite{bertschbertulani}

Theoretically the PDA effect has been studied by several authors.
The sudden approximation, which takes into account the interaction
to all orders but is valid only if the time ordering can be
neglected, provides a simple method to determine the higher-order
effects. Calculations\cite{baur94} show that cross sections tend to
be smaller than those obtained in first- and second-order
perturbation theory, at very small excitation energy which is the
region of interest for nuclear astrophysics. A more advanced
approach which also takes into account the interaction to all orders
is provided by solving directly the time-dependent Schr\"odinger
equation (Refs.\
\cite{esbensen95,bertulani96,kido96,melezhik99,uts99} and references
therein). Here, the motion of the projectile and of the fragments in
the Coulomb field of a heavy target are treated quasi-classically
assuming straight-line trajectories, but the relative motion of the
fragments quantum-mechanically. However, the time evolution in a
Coulomb field does not proceed along a straight line even for large
times because of an additional, logarithmically increasing time
dependence (which is easily seen by integrating the classical
equation of motion in a Coulomb potential). A recent
check\cite{esbbr05} of the assumption that the projectile and the
target do not overlap during the collision (``far-field
approximation'') has shown that it decreases the $S$-factor
extracted from the data of Ref.~\cite{austin2001} by $5\%$. However,
it is not clear how accurate the quasiclassical approach is in the
region where the projectile and the target do overlap.

The CDCC method has also been applied.\cite{mortimer02,matsumoto03}
As usual the effects of the $E2$ transition and higher order
excitations are included. Since the relative motion of projectile
and target, with the huge number of partial waves involved, is
treated quantum-mechanically, such kind of approach becomes
numerically very involved. In some cases\cite{Marta02} the
semiclassical approximation was seen to compare well with
corresponding fully quantal calculations at sub-barrier energies.
This represents a chance for an enormous simplification in the
numerical work since it allows to check and to possibly supersede
the CDCC calculations. At higher energies, where also nuclear
(grazing) collisions can become important, a Glauber approach is
very useful.\cite{BertulaniCG02} Alternatively, in the stationary
framework use has been made of, among others, the
DWBA\cite{banerjee02} and of the quasi-classical
approach.\cite{banerjee93} However, in these investigations the
three-body scattering wave function for the initial (or final) state
is approximated by a product of two two-body or quasi-two-body wave
functions which, as will be explained below, actually completely
eliminates PDA. For instance, in Refs.~\cite{banerjee02,banerjee93}
a ``local'' momentum is introduced which should allow the
incorporation of PDA; but in reality it is a ``global'' momentum
because the distance between the spectator and the other two
particles which are close to each other is rather arbitrarily fixed
at a global value of $10$ fm. Besides, the method used provokes
further questions since the direction of the introduced ``local''
momentum is undetermined.

In a systematic approach, the dissociation amplitude should be
calculated by taking into account the coupling of the different
channels, utilizing, for instance, the FRESCO
code.\cite{austin2001,tostevin2001} However, being presently
interested only in an estimation of the possible effects of
three-body Coulomb final-state interactions which are responsible
for PDA on the breakup cross section at small scattering angles and
relatively high energies, we take resort to some approximations
which are admissible under such conditions. Starting from the prior
form of the dissociation amplitude we introduce for the exact
three-charged particle \wfn occurring therein an approximation
which, while simplifying the calculations, still preserves the
genuine three-body nature. Thus, the PDA effect will be accounted
for in a systematic manner.

The plan of the paper is as follows. Section \ref{NAP} contains a
brief discussion why Coulomb breakup experiments are useful for
obtaining information on capture reactions, and a short summary of
pertaining experiments and their astrophysical relevance. We then
introduce in Sec.\ \ref{gen} the prior-form dissociation amplitude
and the (approximate) \wfn used to describe the three-charged
particle final state. There we also point out that for an
investigation of such breakup processes the post form is
inappropriate for the derivation of physically correct DWBA or
similar approximations. In Sec.\ \ref{results} some cross section
results for the Coulomb breakup reactions
${}^6\tLi+{}^{208}\tPb\longrightarrow \alpha+d+{}^{208}\tPb$ and
${}^8\tB +{}^{208}\textrm{Pb}\longrightarrow {}^7\tBe +
p+{}^{208}\textrm{Pb}$ are presented from which the impact of PDA on
the astrophysical $S$-factor as determined from this reaction can be
estimated. Finally, Sec.\ \ref{summary} contains some conclusions.

We use the units in which $\hbar = c = 1$. Moreover, unit vectors
are denoted by a hat, e.\ g., ${\hat \bx} = \bx/x$.

\section{Coulomb breakup reactions and their relation to problems
of nuclear astrophysics} \protect\label{NAP}

In nuclear astrophysics, radiative capture reactions of the type
\begin{equation}
b + c \to a + \gamma
\label{astro1}
\end{equation}
play a very important role. They can also be studied by the
time-reversed reaction
\begin{equation}
\gamma + a \to b + c \:,
\label{astro2}
\end{equation}
at least in those cases where the nucleus $a$ is in its ground
state. The cross sections of reactions ~(\ref{astro1})
and~(\ref{astro2}) are related by detailed balance,
\begin{equation}
\sigma(b+c \to a +\gamma)=\frac{2(2j_a+1)}{(2j_b+1)(2j_c+1)}
\frac{k_{\gamma}^2}{k_{bc}^2} \sigma(\gamma +a \to b+c\:).
\end{equation}
The relative momentum between particles $b$ and $c$ is given by
$k_{bc}=\sqrt{2\mu_{bc}E_{bc}}$, with $\mu_{bc}$ being the
appropriate reduced mass; the photon momentum is $k_{\gamma}=
(E_{bc}+Q)$, where $Q$ is the $Q$-value of the capture reaction
(\ref{astro1}). Typically one has $k_{\gamma}/k_{bc} \ll 1$ for
energies not in immediate neighborhood of the threshold region
$k_{bc} \approx 0$. Due to this phase space factor the cross section
for the capture reaction (\ref{astro1}) is strongly suppressed in
comparison to that of the dissociation reaction (\ref{astro2}).

In non-resonant charged particle reactions the energy dependence of
the cross section for reaction (\ref{astro1}) is dominated by the
penetration of the Coulomb barrier. This energy dependence is
usually factored out by defining the astrophysical $S$-factor
\begin{equation}
S(E_{bc})= E_{bc} \sigma(b+c \to a +\gamma) \exp(2\pi \eta),
\end{equation}
where $\eta={Z_b Z_c e^2\mu_{bc}}/{k_{bc}}$ is the Coulomb
parameter. It is the most characteristic and important quantity for
nuclear astrophysics.

In fast peripheral collisions of light nuclei with a heavy nucleus
the Coulomb field of the target nucleus creates the equivalent
photon spectrum which allows the extraction of the cross section of
reaction (\ref{astro1}) from Coulomb breakup reactions.\cite{baur86}
Recent developments, especially of radioactive beams, often permit
one to obtain equivalent information with higher-energy beams. The
high-energy Coulomb breakup experiments generally yield higher event
rates; they also yield sometimes information which is not available
in the direct radiative capture reactions.

The ${}^{6}\textrm{Li}$ Coulomb dissociation into $\alpha+d$ has
been a test case of the Coulomb breakup
method.\cite{kiener91,baur96} The experiments were carried out with
a 156 MeV ${}^6$Li beam in Karlsruhe\cite{kiener91}, and with a
60~MeV beam in Heidelberg.\cite{hesselbarth91} This reaction is of
astrophysical importance since the $d(\alpha,\gamma){}^6$Li
radiative capture is the only process by which ${}^6\tLi$ is
produced in the standard primordial nucleosynthesis models.

The radiative capture reaction $t(\alpha,\gamma)^7$Li is
astrophysically relevant for the production of the nuclide $^7$Li in
the early universe and during the nucleosynthesis in stars. The most
accurate direct capture measurement has been performed in
Ref.~\cite{brune94}. The Coulomb dissociation method to investigate
the ${}^7\textrm{Li}\to \alpha + t$ has been applied, too, under
various conditions. We mention here the direct breakup of $^7$Li
scattered at 70~MeV off a $^{120}$Sn target,\cite{shotter84} at 63
MeV off $^{58}$Ni, $^{120}$Sn, $^{144}$Sm, and $^{208}$Pb
targets,\cite{uts90} and at 54~MeV off $^{197}$Au.\cite{gazes92} A
rather detailed measurement was performed recently, with improved
experimental techniques.\cite{tokimoto01}

The reaction ${}^9\mbox{Li} \rightarrow {}^8\mbox{Li} + n$ was
studied with the Coulomb dissociation of a ${}^9\tLi$ beam of 28.53
$A \cdot$\,MeV at MSU\cite{zecher98} to get information on the cross
section for the radiative capture reaction
${}^8\tLi(n,\gamma){}^9\tLi$, which is of importance for the
nucleosynthesis in inhomogeneous big bang models and in Type~II
supernovae.\cite{baur03,Thielemann01} It is also important for
determining the different primordial abundances of Li, Be, B, and C,
in order to limit the degrees of freedom in inhomogeneous big bang
nucleosynthesis.\cite{baur03,Thielemann01}

The $^{14}\mbox{C}(n,\gamma)^{15}\mbox{C}$ reaction plays a role in
neutron-induced CNO cycles of stellar evolution phases beyond the
main sequence. It is also involved in the chain of reactions in
primordial nucleosynthesis in the neutron-rich environment of an
inhomogeneous big bang. Therefore, at MSU a beam of $^{15}$C ions of
$E_{\ti}=35$  $A \cdot$\,MeV was used to study the
${}^{15}\mbox{C}\to {}^{14}\mbox{C}+n$ breakup on C, Al, Zn, Sn, and
Pb targets.\cite{horvath02}

The Coulomb breakup of $^{14}$O was measured at
RIKEN\cite{motobayashi91,serata02} and GANIL\cite{kiener93} in order
to determine the $S$-factor of the $^{13}$N$(p,\gamma) ^{14}$O
capture reaction which is of astrophysical relevance in the hot CNO
cycle.

The reactions $^{11}\mbox{C}(p,\gamma)^{12}\mbox{N}$ which plays a
role in the hot $pp$-chain, and
$^{22}\mbox{Mg}(p,\gamma)^{23}\mbox{Al}$ which is relevant for the
nucleosynthesis of $^{22}$Na in Ne-rich novae, were studied at RIKEN
by the Coulomb dissociation
method.\cite{motobayashi02,minemura02,gomi02} Their
results\cite{motobayashi02,minemura02} are consistent with the GANIL
measurement.\cite{lef95} The accuracy of the value for the radiative
width of the 1.19~MeV level in $^{12}$N was improved.

Finally, the ${}^{7}$Be(p,$\gamma$)$^{8}$B radiative capture
reaction is important for the solar neutrino problem which has been
solved recently after 40 years of discussions and measurements by
many groups. It determines the production rate of ${}^{8}$B which is
the source of high-energy electron neutrinos $\nu_e$. In fact, the
astrophysical $S$-factor for this reaction must be known to $\pm
5\%$ in order that this uncertainty not be the dominant error in
predictions of the solar $\nu_e$ flux. The Kamiokande\cite{fukuda02}
and SNO\cite{ahmad01,aliani02} measurements of the neutrino flux
from the sun confirmed the standard solar model (cf.\
Ref.~\cite{bahcall01}). Also the Coulomb dissociation
$^{8}\textrm{B}\to {}^{7}\textrm{Be}+p$ has been studied at
different radioactive beam facilities under different kinematic
conditions and energies (46.5, 51.9, 83, 90, and 254 $A \cdot$\,MeV)
at RIKEN,\cite{motob94,kikuchi98} MSU,\cite{austin2001} and
GSI.\cite{iwasa99}

\section{General Coulomb dissociation amplitude}\protect\label{gen}

Since in the present investigation the main emphasis lies on the
elaboration of possible effects of three-body Coulomb final-state
interactions on the breakup cross section at small scattering angles
and relatively high energies, we take resort to some approximations
which are admissible under such conditions.

As starting point we choose the exact dissociation amplitude for the
reaction (\ref{brkp1}) in the prior form,
\begin{equation}
M_{\tif}^{\tprior} = \langle \Psi_{\tf}^{(-)} | \Delta V |\,
\varphi_{a} \Phi_{\ti}^{(+)} \rangle,\; \Delta V= V_{bA} + V_{cA} -
U. \label{rctam1}
\end{equation}
Here, $V_{jA}= V_{jA}^{\tN} + V_{jA}^{\tC}$ is the interaction
potential of particle $j(=b,c)$ with the target particle $A$
containing nuclear ($V_{jA}^{\tN}$) and Coulombic ($V_{jA}^{\tC}$)
parts. The distorted wave $\Phi_{\ti}^{(+)}$ in the initial channel
is the continuum solution of the two-particle Schr\"odinger equation
for the full optical potential $U=U^{\tC} + U^{\tN}$ and describes
the scattering of particles $a$ and $A$ with relative kinetic energy
$E_{\ti}$; $\varphi_{a}$ is the $ a=(bc)$ bound state wave function
with binding energy $\varepsilon_{bc}$, and $\Psi_{\tf}^{(-)}$ the
full final-state wave function for three charged particles $b, c,$
and $A$ in the continuum, with the total energy
\begin{equation}
E \equiv E_{\tf} + E_{bc} \equiv E_{\ti} - \varepsilon_{bc}.
\label{eneqn1}
\end{equation}
The various kinetic energies are defined as $E_{bc} =
{k^{2}_{bc}}/{2 \mu_{bc}}$ and $E_{\text{i,f}} =
{q_{\text{i,f}}^{2}}/{ 2 \mu_{aA}};\;\bk_{bc}$ is the relative
momentum of particles $b$ and $c$ in the exit channel,
$\bq_{\text{i,f}}$ the relative momentum of $a$ and $A$ in the
entry, and of the \cm of $(b+c)$ and $A$ in the exit channel,
respectively. The reduced mass of particles $i$ and $j$ is denoted
by $\mu_{ij}= m_{i} m_{j}/(m_{i}+m_{j})$.

We assume that the incident center-of-mass energy $E_{\ti}$ of the
projectile $a$ is high enough and that the scattering angle $\theta$
of the \cm of the pair $(b+c)$ is so small (at given $E_{\ti}$) that
the impact parameter is much larger than the nuclear interaction
radius $R_{\tN}$ between $a$ and $A$. Such conditions are typical
for Coulomb breakup processes. Then all potentials in (\ref{rctam1})
can be approximated by their Coulombic parts, $\Delta V \approx
\Delta V^{\tC} = V_{bA}^{\tC} + V_{cA}^{\tC} - U^{\tC}.$ Note that
under such conditions, $\Phi_{\ti}^{(+)}$ is just a two-body Coulomb
scattering wave function for the potential $U^{\tC}$.

The dissociation amplitude (\ref{rctam1}) requires knowledge of the
three-charged particle wave function $\Psi_{\tf}^{(-)}$ which is a
solution of the three-body Schr\"odinger equation with incoming-wave
three-body boundary conditions,
\begin{equation}\label{schrod}
(E-T_{\br}-T_{\brho}- V_{bc} - V_{bA} - V_{cA})
\Psi^{(-)}_{\bk_{bc},\bq_{\tf}}(\br,\brho)=0.
\end{equation}
Here, $\br=\br_{b}-\br_{c}$ and $\brho=(m_{b}\br_{c} +
m_{c}\br_{b})/(m_{b}+m_c) - \br_{A}$, where $\br_{j}$ is the
coordinate vector of particle $j$, while $\bk_{bc}$ and $\bq_{\tf}$
are the momenta canonically conjugated to $\br$ and $\brho$,
respectively. Hence, $T_{\br}$ and $T_{\brho}$ are the kinetic
energy operators corresponding to the relative motion of $b$ and
$c$, and to the relative motion of the \cm of the pair $(b+c)$ and
the target $A$, respectively.

Since such a \wfn is neither known and nor numerically calculable at
present, physically justifiable approximations are unavoidable. Such
can be found by the following consideration. The amplitude
(\ref{rctam1}) contains the overlap of the initial bound state wave
function $\varphi_{a}(\br)$ for the bound pair $(bc)$ and the
three-body $(b + c + A)$ final-state continuum wave function. The
former limits the range of variation of $r$. But the range of $\rho$
is not confined and, hence, extends to infinity. Thus, if $\rho \gg
R_{\tN}'$, where $R_{\tN}'$ is the larger of the nuclear interaction
radii between particles $b$ or $c$ and $A$, it suffices to find
solutions of the \SE restricted to the asymptotic domain
$\Omega_{A}: \{ r\ll\rho,\; R_{\tN}' \ll \rho \}$, \ie where the
distance $r$ between the fragments $b$ and $c$ is much smaller than
that between their center of mass and $A$.

It is easy to see that in $\Omega_{A}$ the \SE (\ref{schrod})
reduces to
\begin{equation}
\left\{E - T_{\br} - T_{\brho} - V_{bc}(r) - U^{\tC}(\rho)\right\}
\Psi^{(+)}_{\bk_{bc},\bq_{\tf}} (\br,{ \brho}) = O(r^2/{\rho}^{2}),
\label{eq2}
\end{equation}
since there
\begin{equation}
V_{bA}(\br_{b}) + V_{cA}(\br_{c}) \stackrel{\Omega_{A}}{\approx}
V_{bA}^{\tC}(\br_{b}) + V_{cA}^{\tC}(\br_{c}) = U^{\tC}(\brho)+
O(r^2/\rho^2), \label{voa}
\end{equation}
with
\begin{equation}
V_{jA}^{\tC}(\br_{b})= {Z_{j} Z_{A} e^{2}}/{|{\brho} +
\varepsilon_{j}\mu_{bc} \br /m_{j}|}, \; U^{\tC}(\rho)= {Z_{a} Z_{A}
e^{2}}/{\rho},\; j=b,c. \label{coptpot1}
\end{equation}
Here, $\varepsilon_{j}= +1 (-1)$ for $j=b(c)$, and $Z_{j}e$ is the
charge of particle $j$.

Since the Hamiltonian in Eq. (\ref{eq2}) separates in the variables
$\br$ and $\brho$, the obvious solution is provided by a product of
two two-body scattering wave functions which we write for later
purposes as
\begin{equation}
\Psi_{\tf}^{(-)}(\br, \brho) \equiv \Psi^{(-)}_{\bk_{bc}, \bq_{\tf}}
(\br,\brho) \approx e^{ i \bk_{bc} \cdot \br + i\bq_{\tf} \cdot
\brho}\; \tilde \psi_{\bk_{bc}}^{(-)} (\br ) \tilde
\Psi^{(-)}_{\bq_{\tf}} (\brho). \label{fnstconv1}
\end{equation}
Here, $\psi_{\bk_{bc}}^{(-)}(\br) \equiv e^{ i \bk_{bc} \cdot \br }
\tilde \psi_{\bk_{bc}}^{(-)} (\br ) $ describes the relative motion
of particles $b$ and $c$, interacting via the potential $V_{bc}(r)$;
and $\Psi_{ \bq_{\tf}}^{(-)} (\brho)\equiv e^{ i\bq_{\tf} \cdot
\brho}\; \tilde \Psi^{(-)}_{\bq_{\tf}} (\brho)$ is the scattering
wave function of the \cm of $(b+c)$ and $A$, interacting via the
optical potential $U^{\tC}(\rho)$. When inserted in the exact
expression (\ref{rctam1}), this leads to the DWBA amplitude
conventionally employed in analyses of Coulomb breakup reactions,
\be M_{\tif}^{\tDWBA} = \langle \psi_{\bk_{bc}}^{(-)}
\Psi^{(-)}_{\bq_{\tf}} | \Delta V^{\tC} |\, \varphi_{a}
\Phi_{\ti}^{(+)} \rangle. \label{dwba}\ee We mention that frequently
a wave function $\Psi_{ \bq_{\tf}}^{(-)} (\brho)$ calculated with
the full optical potential $U^{\tC}(\rho) + U^{\tN}(\rho)$ is used
here.\cite{baur84,bertulani91}

Approximation (\ref{fnstconv1}) which has dramatic consequences is
motivated by two arguments: (i) the right-hand side of Eq.
(\ref{fnstconv1}) is indeed an exact solution of the \SE (\ref{eq2})
restricted to the domain $\Omega_{A}$ which is relevant for Coulomb
breakup processes of interest here, and (ii) its use greatly
simplifies the evaluation of the dissociation amplitude
(\ref{rctam1}).

While the second argument obviously applies (see below), the first
one is grossly misleading. Namely, it comes as a surprise that the
condition that Eq.\ (\ref{eq2}) be satisfied up to $O(r^2/{
\rho^{2}})$ order {\em does not suffice to provide a unique
solution}.\cite{alt93} To remove this non-uniqueness and to select
the physically acceptable solution, an additional constraint has to
be imposed. Such is provided by the requirement that the {\em
correct} solution of the \SE (\ref{eq2}) in $\Omega_{A}$ has to
smoothly match with the solution of the exact \SE (\ref{schrod}) in
the asymptotic region $\Omega_{0}$ where all three particles $b, c,$
and $A$, are well separated, \ie $r\to \infty,\; \rho \to \infty$,
since $ \Omega_{0} \cap \Omega_{A} \neq 0$. The leading term of the
asymptotic solution in the asymptotic domain $\Omega_{0}$ is given
by the Redmond three-body distorted wave,\cite{red73}
\begin{eqnarray}
\Psi^{(-)}_{ \bk_{bc},\bq_{\tf}}(\br,\brho) \; &\stackrel{\Omega
_{0}}{\approx} \; &e^{i \bk_{bc} \cdot \br + i\bq_{\tf} \cdot
\brho}\; e^{-i \eta_{bc} \ln\zeta_{bc}}\,e^{-i \eta_{bA}
\ln\zeta_{bA}}
\,e^{-i \eta_{cA} \ln\zeta_{cA}} \label{rmwfnas}\\
&\stackrel{\Omega _{0}}{\approx}& \; e^{ i \bk_{bc} \cdot \br +
i\bq_{\tf} \cdot \brho}\; \tilde{\psi}_{ {\rm {\bf
{k}}}_{bc}}^{C(-)}({\rm {\bf {r}}}_{bc })\tilde{\psi}_{ {\rm {\bf
{k}}}_{bA}}^{C(-)}({\rm {\bf {r}}}_{bA })\tilde{\psi}_{ {\rm {\bf
{k}}}_{cA}}^{C(-)}({\rm {\bf {r}}}_{cA }), \label{rmwfn}
\end{eqnarray}
where the second form\cite{mer77} is asymptotically equivalent to
the first one. Here, $\eta_{ij}$ is the Coulomb parameter,
$\,\zeta_{ij}= k_{ij}\,r_{ij} - {\rm {\bf k}}_{ij} \cdot {\rm {\bf
r}}_{ij}$, and ${\rm {\bf r}}_{ij} ({\rm {\bf k}}_{ij})$ the
relative coordinate (relative momentum) between particles $i$ and
$j$. The wave functions $ \psi^{C(-)}_{\bk_{ij}}(\br_{ij}) \equiv
e^{ i \bk_{ij} \cdot \br_{ij} } \tilde \psi^{C(-)}_{\bk_{ij}}
(\br_{ij})$ describe the scattering of particles $i$ and $j$,
interacting via the Coulomb potential $V^{C}_{ij}(r_{ij})$ (with
incoming wave boundary condition).

It is apparent, and in fact a simple exercise to check, that
expression (\ref{fnstconv1}) does not match in the leading order
with the asymptotic solution (\ref{rmwfnas}) in $\Omega_{0}$ and,
hence, is {\em not admissible}. That defect notwithstanding, it is
conventionally taken as the approximation for the final-state
three-body wave function $\Psi_{\tf}^{(-)}$.

Thus, an alternative approximation for $\Psi_{\tf}^{(-)}$ must be
found which, while satisfying the above constraints, nevertheless
renders the dissociation amplitude calculable. A three-body \wfn
which satisfies all these conditions has been derived in
Refs.~\cite{alt93,muk96}. The leading order term\cite{alt93} can be
written as
\begin{eqnarray}
\Psi^{(-)(\tas)}_{\tf}(\br, \brho) \; &\stackrel{\Omega
_{A}}{\approx} \; &e^{i \bk_{bc} \cdot \br + i\bq_{\tf} \cdot
\brho}\; \tilde \psi^{(-)}_ {\bk_{bc}(\brho)} (\br) \,e^{-i
\eta_{bA} \ln\zeta_{bA}}
\,e^{-i \eta_{cA} \ln\zeta_{cA}} \nonumber\\
&\simeq& e^{ i \bk_{bc} \cdot \br + i\bq_{\tf} \cdot \brho}\; \tilde
\psi^{(-)}_ {\bk_{bc}(\brho)} (\br) \tilde \psi_{\bk_{bA}}^{(-)}
(\br_{bA} ) \tilde \psi_{\bk_{cA}}^{(-)} (\br_{cA} ) . \label{aswf1}
\end{eqnarray}
The new wave function $\psi^{(-)}_{\bk_{bc}(\brho)}(\br) \equiv e^{
i \bk_{bc}(\brho)\cdot \br_{bc}} \tilde \psi^{(-)}_{\bk_{bc}(\brho)}
(\br)$ describes the scattering of particles $b$ and $c$,
interacting with the full potential $V_{bc}(r)$ and moving with
relative momentum $\bk_{bc}(\brho)$. As a consequence of the
long-range Coulombic three-body correlations between $b$ and $A$ and
between $c$ and $A$ the latter, however, depends on the distance of
their \cm to the charged particle $A$, and is given by
\begin{eqnarray}
\bk_{bc}(\brho)={\bf k}_{bc}- \frac{\mathbf{a}(\hrho)}{\rho}, \;
{\mathbf{a}(\hrho)} :=\mu_{bc}\left\lbrack\frac{\eta_{b
A}}{m_b}\frac{(\hk_{b A}+ \hrho)}{(1 + \hrho \cdot \hk_{b A})}-
\frac{\eta_{c A}}{m_c} \frac{(\hk_{c A}+ \hrho)}{(1 + \hrho \cdot
\hk_{c A})} \right \rbrack. \label{krho}
\end{eqnarray}
It is not difficult to see that the \wfn (\ref{aswf1}) has the
desired properties: (i) it is asymptotically solution of the
asymptotic \SE (\ref{eq2}) and (ii) it smoothly matches with the
Redmond \wfn in $\Omega_{0}$ and, hence, is asymptotically also
solution of the original \SE (\ref{schrod}).

It is apparent that the three-charged particle wave function
(\ref{aswf1}) which, as we emphasize once more, is the only
physically admissible, asymptotically exact solution of the
three-body \SE (\ref{eq2}) for use in the dissociation amplitude
(\ref{rctam1}), is much more complicated than (the asymptotic part
of) the simple factorized form (\ref{fnstconv1}) conventionally used
in DWBA calculations. Insertion into (\ref{rctam1}) yields for the
appropriate dissociation amplitude
\begin{equation}
M_{\tif}^{\tprior} \approx \langle \Psi_{\tf}^{(-)(\tas)} | \Delta
V^{\tC} |\, \varphi_{a} \Phi_{\ti}^{(+)} \rangle.
\label{dissamp}
\end{equation}

Since the integration over $r$ is limited by the bound state wave
function $\varphi_{a}(\br)$ of the projectile $a$, and if the impact
parameter is larger than the nuclear interaction radius $R_{\tN}$
between $a$ and $A$, we can utilize the multipole expansion for
$\Delta V$. It leads to the multipole expansion of the dissociation
amplitude (\ref{dissamp}):
\begin{equation}
M_{\tif}=\sum_{LM}\frac{4 \pi}{2 L + 1} M_{LM}. \label{mlexp}
\end{equation}
The multipole transition matrix elements $M_{LM}$ are defined as
\begin{eqnarray}
M_{LM} &=& Z_A Z^{(\teff)}_L e^{2} \int d \brho \int
d\br\, \Psi_{\tf}^{(-)(\tas)*}(\br, \brho) \nonumber \\
&&\times \frac{r^{L}}{\rho^{L+1}} \, Y_{LM}^{*}(\br) \,
Y_{LM}(\hrho) \, \varphi_{a}(\br) \, \Phi_{\bq_{\ti}}^{(+)}(\brho),
\label{rctam2}
\end{eqnarray}
with $Z^{(\teff)}_L=\mu_{bc}^{L} \left[(-1)^{L} {Z_{b}}/{m_{b}^{L}}
+ { Z_{c}}/{m_{c}^{L}} \right]$ and $\Phi_{\ti}^{(+)}(\brho) \equiv
\Phi_{\bq_{\ti}}^{(+)}(\brho)$. Note that if in
$\Psi_{\tf}^{(-)(\tas)}$ also the next-to-leading-order terms are
taken into account, two more terms of similar structure but
suppressed by an additional factor $1/\rho$ occur in
(\ref{rctam2}).\cite{muk96}

The right-hand side of Eq.\ (\ref{rctam2}) contains a
six-dimensional quadrature. This is to be contrasted with the
conventional approach. There, the multipole matrix elements of the
DWBA amplitude (\ref{dwba}) separate into a product of two terms,
each requiring three-dimensional quadratures. This greatly reduces
the numerical expense.

Use of the amplitudes (\ref{dissamp}), resp.\ (\ref{rctam2}), which
are based on the prior-form dissociation amplitude has two crucial
consequences. First, since the \wfn occurring therein has the
correct asymptotic behavior this expression is consistent with the
general principles of scattering theory of charged particles. This
is to be contrasted with the standard DWBA of the dissociation
amplitude (\ref{dwba}) the derivation of which usually starts from
the post form, $ M_{\tif}^{\tpost} = \langle \psi_{bc}^{(-)}
\Phi_{aA}^{(-)} | \Delta V |\, \Psi_{\ti}^{(+)} \rangle,$ with $
\Delta V$ as in Eq.\ (\ref{rctam1}). The final-state wave function
is given by the right-hand side of Eq.\ (\ref{fnstconv1}). The exact
initial-state three-body \wfn is then approximated by the product
$\Psi_{\ti}^{(+)}\approx\varphi_{a} \Phi_{\ti}^{(+)}$. However, in
the presence of Coulomb potentials acting between all three
particles, {\em the post-form amplitude is not valid, in contrast to
the prior form (\ref{rctam1}), and hence is not equivalent to the
latter}. The reason is that the final-state product wave function
occurring therein does not produce the correct behavior in any
asymptotic region of configuration space. Secondly, as mentioned
earlier, the PDA effect occurs for $Z_{b}/m_{b} \not= Z_{c}/m_{c}$,
since it is caused by the difference in the Coulomb forces exerted
by the target on each fragment $b$ and $c$, which leads to the
different accelerations. This fact gives rise to a change of the
relative velocity of the fragments when they finally have reached
the detectors as compared to their relative velocity at the moment
of the decay of the projectile $a$. Use of approximation
(\ref{fnstconv1}), however, completely eliminates the PDA effect
because (i) the wave function part describing the scattering of the
fragments $b$ and $c$ in the Coulomb field of the target $A$ is
approximated by $\Psi_{\bq_{\tf}}^{(-)}(\brho)$ which describes the
scattering of the \cm of the pair $(b+c)$ off the target rather than
the scattering of the individual fragments, and, related to it, (ii)
the relative motion wave function $\psi_{\bk_{bc}}^{(-)}(\br)$ of
the fragments $b$ and $c$ in the continuum contains the asymptotic
relative momentum between $b$ and $c$ and, hence, does not allow for
its modification due to the action of the target Coulomb field
between the breakup point and the counter.

\section{Applications}\protect\label{results}

\subsection{${}^6\!\tLi$ breakup into $\alpha$-particle plus deuteron}

As a first application of the proposed method we have
computed\cite{irg97} the triple-differential cross section (TDCS)
for the reaction
\begin{equation}\label{6Li}
{}^6\tLi+{}^{208}\tPb\longrightarrow \alpha+d+{}^{208}\tPb,
\end{equation}
for a projectile energy $E_{\ti}=156$ MeV and scattering angles
$\theta_\alpha= \theta_d=3^\circ$. In this investigation only the
leading term of the asymptotic \wfn (\ref{aswf1}) has been retained.
Moreover, because of the peripherality of the reaction it suffices
to take into account only the Coulomb interaction between the
fragments in the final state so that $\psi^{(-)}_{{\bk}_{\alpha
d}(\brho)} (\br)$ is simply a Coulomb scattering wave function,
albeit with local momentum $\bk_{\alpha d}(\brho)$. Since for $\rho
> R_{\tN}$, the additional term ${\mathbf{a}(\hrho)}/{\rho}$ in the
local momentum turned out to be much smaller in magnitude than the
asymptotic momentum $\bk_{\alpha d}$, this \wfn was expanded in
powers of $|\mathbf{a}(\brho)|/\rho k_{\alpha d}$. Keeping only the
leading terms we arrive at ($\psi(z)$ the digamma function)
\bay\label{psi-exp} \psi^{(-)}_ {\bk_{\alpha d}(\brho)} (\br)
&\approx& \left[1+\frac{A(\hrho,\bk_{\alpha
d})}{\rho}\right] \psi^{C(-)}_ {\bk_{\alpha d}}(\br),\\
A(\hrho,\bk_{\alpha d})&=&\frac{\ba(\hrho)\cdot\hk_{\alpha
d}\,\eta_{\alpha d}}{k_{\alpha d}}
\left[\frac{\pi}{2}+i\psi(1-i\eta_{\alpha d})\right]. \eay Use of
(\ref{psi-exp}) greatly simplifies the calculation of the amplitudes
$M_{LM}$ since each of them separates into a product of two
quantities contaning integration over either $\br$ or $\brho$ only.
Thus, in this approximation the whole effect of PDA resides in the
factor $A(\hrho,\bk_{\alpha d})$; or explicitly, for
$A(\hrho,\bk_{\alpha d})=0$ PDA does not occur.

Both $E1$ and $E2$ transitions were included. In the isospin
formalism, for the ${}^6\tLi\to\alpha+d$ breakup the $E1$ transition
is forbidden ($\Delta T=0$); thus the main contribution comes from
$E2$. Nevertheless, since the isospin rule is violated and the
charge-over-mass ratio $Z_\alpha/m_\alpha$ of the $\alpha$-particle
differs only little from that of the deuteron, $E1$ contributes
comparably to $E2$ for $E_{\alpha d}<100$ keV (see also
Ref.~\cite{zhus90} where the contribution of $E1$ to the reaction
$\alpha+d\to {}^6\tLi+\gamma$ is discussed).

\begin{figure}[h]
\centerline{\psfig{file=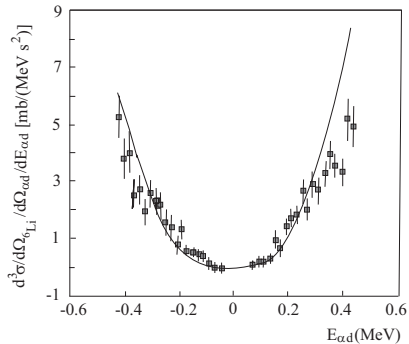,width=2.1in}} \vspace*{8pt}
\caption{The TDCS for the reaction
${}^6\tLi+{}^{208}\tPb\longrightarrow \alpha+d+{}^{208}\tPb$ at a
projectile energy $E_{\ti}= 156$ MeV and scattering angle
$\theta=3^\circ$ of the fragments, as a function of the $\alpha-d$
relative kinetic energy $E_{\alpha d}$. Negative (positive) values
correspond to $\alpha-d$ relative velocities ${\it v}_{\alpha d} < 0
\,\,({\it v}_{\alpha d}> 0)$. Experimental data are from
Ref.~\protect\cite{kiener91}. } \label{fig:5}
\end{figure}

In Fig.~\ref{fig:5} we present the calculated TDCS. Inspection shows
that for $|E_{\alpha d}| < 300$ keV the agreement between the
experimental\cite{kiener91} and theoretical results is very good.
Hereby, the notation is such that positive (negative) values of
$E_{\alpha d}$ are associated with ``forward'' and ``backward''
emission of the $\alpha$-particle in the \cm system of the decaying
${}^6$Li nucleus. The astrophysical $S_{24}$-factor for $E_{\alpha
d}=0$ extracted therefrom (by linear extrapolation into the energy
range $E_{\alpha d}<20$ keV) is $S_{24}(0) = 1.2$ nb$\cdot$MeV. This
value agrees with that obtained\cite{akr95} from a calculation of
the cross section for the radiative capture
$d(\alpha,\gamma){}^6\textrm{Li}$ at ultra-low energies $E_{\alpha
d}$, but disagrees with the old extrapolated value
$S_{24}^{\texp}(0)=9.1\pm1.8$ nb$\cdot$MeV in Ref.~\cite{kiener91}.
Note that an earlier analysis\cite{bw89} of the same data gave a
value of $S_{24}(0)=1.7$ nb$\cdot$MeV.

\begin{figure}[ht]
\centerline{\psfig{file=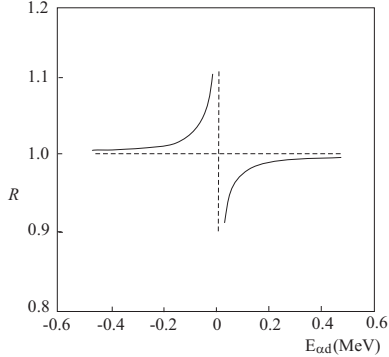,width=2.0in}} \vspace*{8pt}
\caption{Ratio $ \frac{d^3\sigma(A=0)/d\Omega_{{}^6Li} /
d\Omega_{\alpha d}/dE_{\alpha d}} {d^3\sigma(A\neq
0)/d\Omega_{{}^6Li}/d\Omega_{\alpha d}/dE_{\alpha d}}$ of the TDCS
calculated without ($A(\hrho,\bk_{\alpha d})=0$) and with
($A(\hrho,\bk_{\alpha d})\neq 0$) allowance for three-body Coulomb
effects, as function of $\alpha-d$ relative kinetic energy
$E_{\alpha d}$. Negative and positive values of $E_{\alpha d}$ have
the same meaning as in Fig.~\ref{fig:5}. } \label{fig:6}
\end{figure}

Fig.~\ref{fig:6} illustrates the role of the factor
$A(\hrho,\bk_{\alpha d})$,
which comprises within our approximations all three-body Coulomb
final-state interaction effects to the cross section, in the energy
region $|E_{\alpha d}| > 20$ keV (because of our assumption
$|A(\hrho,\bk_{\alpha d})|/\rho \ll 1$ the curves cannot be
extrapolated to the point $E_{\alpha d}= 0$). Nevertheless it is
apparent that the ratio of the cross sections calculated with the
conventional asymptotic \wfn $[A(\hrho,\bk_{\alpha d})=0]$ to the
one calculated with the asymptotically correct wave function, which
takes into account three-body Coulomb effects $[A(\hrho,\bk_{\alpha
d})\neq 0]$, behaves differently in the regions ${\it v_{\alpha d}}
< 0$ and ${\it v_{\alpha d}} > 0$. That is, three-body Coulomb
effects, together with the interference of $E1$ and $E2$
transitions, give rise to an asymmetry of the cross section ratio
with respect to the point ${\it v_{\alpha d}} = 0$. Hence, it is
compulsory to accurately take them into account when deducing the
$S_{24}$-factor from the TDCS.

\subsection{${}^8\!\tB$ breakup into ${}^7\!\tBe$ plus proton}

As a further example we have investigated\cite{aim03,alt05} the
breakup reaction
\begin{equation}\label{8B}
{}^8\tB +{}^{208}\textrm{Pb}\longrightarrow {}^7\tBe +
p+{}^{208}\textrm{Pb},
\end{equation}
at \B beam energies $E_{\ti}=46.5$ $A \cdot$\,MeV\cite{motob94} and
$E_{\ti}=83$ $A \cdot$\,MeV,\cite{austin2001} taking into account
dipole and quadrupole contributions. \Bep relative kinetic energies
are varied between 100 keV and 1 MeV. Recall that this reaction
allows the extraction of the astrophysical $S$-factor for the
important radiative capture reaction ${}^7\textrm{Be} + p\to
{}^8\textrm{B}+\gamma$. The final \Be nucleus is assumed to be in
the ground state. Note that in the kinematic region of interest,
both the binding energy $\varepsilon_{\tBep}=0.137$ MeV of the
proton to the \Be nucleus as well as the kinetic energy of the
relative motion of the proton and \Be in the final state are much
smaller than the collision energy $E_{\ti}$.

When using the asymptotic three-body wave function including
next-to-leading-order terms\cite{muk96} for the final state, the
breakup amplitude is given as a sum of three terms. The first one
arises from the leading term (\ref{dissamp}) (of $O(\rho^0)$) and is
expected to give the largest contribution. Indeed, it turned out
that the cross section is completely governed by it. The other terms
are suppressed by an additional factor $1/\rho$ in the wave
function, contributing less than 1\% to the cross section.

The dipole transition to the \Bep $s$-wave continuum state is found
to dominate the amplitude; the contribution of the dipole transition
to the $d$-wave state is small at small relative kinetic energies
$E_{\tBep}$ but increases up to 30\% at 1 MeV. Moreover, in our
approach the quadrupole amplitude depends only on the projectile
energy $E_{\ti}$ and on the \Bep relative energy $E_{\tBep}$ but not
on the scattering angle $\theta$. Its contribution relative to that
coming from the dipole amplitude changes from a few per cent at
small $E_{\tBep}$ and small $\theta$ to up to 30\% near the grazing
angle. We mention that the same behavior of the quadrupole term was
observed in a calculation using first- and second-order perturbation
theory in the semiclassical approach.\cite{baur94} The consequence
is that for radiative capture reactions at energies of interest in
nuclear astrophysics, the quadrupole amplitude is negligible. For
details we refer to Ref.~\cite{alt05}.

Typical results are presented in Tables \ref{table1} and
\ref{table3}. Table \ref{table1} shows the ratio of the double
differential cross section (DDCS), calculated with the (asymptotic)
three-body wave function (\ref{aswf1}), as compared to the one using
the conventional ansatz (\ref{fnstconv1}) for the final state, at
incident energies $E_{\ti}=46.5$ $A \cdot$\,MeV.
\begin{table}[ht]
\tbl{Ratio $R(\theta)$ of the DDCS for the reaction \Pbr, calculated
with the asymptotic three-body Coulomb wave function
(\protect\ref{aswf1}), to the DDCS calculated using the conventional
wave function (\protect\ref{fnstconv1}), as a function of the
scattering angle $\theta$, for various values of the \Bep relative
energy. The projectile energy is $E_{\ti}=46.5$ $A \cdot$\,MeV.
\label{table1}} {\begin{tabular}{@{}cccccccccccc@{}} \toprule
$E_{\tBep}$ & \multicolumn{11}{c}{$R(\theta)$} \\
\cline{2-12}
[MeV]&$1.0^\circ$&$1.5^\circ$&$2.0^\circ$&$2.5^\circ$&$3.0^\circ$&$3.5^\circ$&
 $4.0^\circ$&$4.5^\circ$&$5.0^\circ$&$5.5^\circ$&$6.0^\circ$\\
\cline{1-12}
 0.1&1.16&1.26&1.37&1.49&1.63&1.77&1.91&2.04&2.17&2.30&2.41\\
\cline{1-12}
 0.15&1.02&1.04&1.07&1.11&1.14&1.16&1.19&1.21&1.23&1.24&1.25\\
\cline{1-12}
 0.2&1.01&1.01&1.02&1.03&1.03&1.03&1.03&1.03&1.02&1.02&1.01\\
\cline{1-12}
 0.4&1.00&1.00&0.99&0.99&0.98&0.98&0.97&0.96&0.95&0.94&0.92\\
\cline{1-12}
 0.6&1.00&1.00&1.00&0.99&0.99&0.99&0.98&0.98&0.98&0.97&0.97\\
\cline{1-12}
 0.8&1.00&1.00&1.00&1.00&1.00&1.00&0.99&0.99&0.99&0.99&0.99\\
\cline{1-12}
 1.0&1.00&1.00&1.00&1.00&1.00&1.00&1.00&1.00&1.00&1.01&1.01\\
 \botrule
\end{tabular}}
\end{table}
The former takes into account three-body Coulomb final state
interactions and, hence, PDA while the latter does not. Thus,
deviation of this ratio from unity is a direct measure of the
influence of PDA. We only mention that a calculation at 83 $A
\cdot$\,MeV shows that PDA effects decrease with increasing
projectile energy and relative energy of the outgoing fragments. But
at small relative energies and large scattering angles, PDA is
important, the deviation from unity reaching more than 10\%.
\begin{table}
\tbl{The ratio as in Table \protect\ref{table1} but for the SDCS
$d\sigma/dE_{\tBep}$, at $E_{\ti} = 46.5$ and $83$ $A \cdot$\,MeV,
obtained by integrating the DDCS over the scattering angle $\theta
\in [2^{\circ}-6^{\circ}] \mbox{ and } [0.1 - 1.77^{\circ}] $,
respectively.
\label{table3}} {\begin{tabular}{@{}lccccccc@{}}
\toprule
$E_{\tBep}$ [MeV]&0.1 &0.15 &0.2 &0.4 &0.6 &0.8 &1.0 \\
\hline Ratio ($E_{\ti} = 46.5$ $A \cdot$\,MeV)&1.90&1.18& 1.03&
0.97& 0.98&1.00& 1.00\\
Ratio ($E_{\ti} = 83$ $A \cdot$\,MeV)&1.13&1.02&1.00 & 1.00 & 1.00 &
1.00 & 1.00 \\
\botrule
\end{tabular}}
\end{table}
Moreover, from Table \ref{table3} we conclude that PDA affects the
single differential cross section (SDCS) less than the DDCS. This is
understandable because PDA is large near the grazing angle where the
cross section is small; and the integration over the scattering
angle of the DDCS levels out its contribution. We mention that our
approach may produce a large error near the grazing angle where the
nuclear interaction between the projectile and the target becomes
important which has been omitted here.

From these results the following conclusions can be drawn.\\
(i) At 46.5 $A \cdot$\,MeV incident energy, the DDCS is enhanced by
three-body Coulomb final-state interactions only at the lowest
relative energies $E_{\tBep}$ while in the medium energy range it is
reduced; at the highest energies considered their effect becomes
negligible. That is, the three-body effects increase the probability
of dissociation of \B up to energies of
$E_{\tBep}\sim 200$ keV, but start suppressing it at $\sim 400$ keV. \\
(ii) Three-body Coulomb effects become more important as $\theta$
increases. The rather small influence at very small $\theta$ is due
to the fact that there large impact parameters dominate for which
the three-body correlations are weak. Thus, in order to avoid PDA
effects, experiments at $E_{\ti}=46.5$ $A \cdot$\,MeV should be
performed at $\theta \leq 4^{\circ}$. \\
(iii) On the other hand, PDA effects decrease as the incident energy
increases, cf.\ Table \ref{table3} which compares the SDCS ratios
with and without three-body effects, at $E_{\ti}=46.5$ $A
\cdot$\,MeV\cite{motob94} and $83$ $A \cdot$\,MeV.\cite{austin2001}
Thus, PDA effects can also be made negligibly small by choosing
sufficiently large projectile energies.\\
(iv) Since our present approach is based on an asymptotic
three-particle wave function and, hence, does not allow to perform
the calculations for small enough \Bep relative kinetic energies,
extrapolation to $E_{\tBep}=0$ seems too arguable. Hence, no
reliable quantitative predictions about the influence of three-body
Coulomb effects on the Coulomb dissociation cross section and the
$S_{17}$-factor to be extracted from it can be made. However,
qualitative conclusions, in particular under which kinematic
conditions PDA effects are negligible, can already be drawn. This
shortcoming is shared also by the standard DWBA approaches based on
Eq.\ (\ref{dwba}) since there also only a solution, and indeed an
unacceptable one, of the asymptotic \SE (\ref{eq2}), and not a
genuinely calculated three-body \wfn is used.

\section{Summary and outlook}
\protect\label{summary}

Peripheral collisions of medium- and high-energy nuclei, that is
collisions in which these nuclei pass a heavy, fully stripped target
nucleus at distances beyond the nuclear interaction radius and which
are therefore dominated by electromagnetic interactions, have
evolved into an important tool of the nuclear astrophysical
research, because of the intense source of quasi-real photons
available from the target in such collisions. In addition new
experimental facilities, in particular the present and forthcoming
radioactive beam factories, render feasible the efficient
investigation of photo-interactions with (stable and unstable)
nuclei, such as single- and multi-photon excitations and
electromagnetic dissociation.

In Coulomb breakup processes, among the various so-called
higher-order effects the post-decay acceleration (PDA) of the
fragments in the target Coulomb field which is due to three-body
Coulomb final-state interactions plays an important role, especially
at low and medium collision energies. Various theoretical and
computational methods presently in use for taking them into account
have been briefly sketched. We have, in particular, emphasized that
the frequently used conventional separation of the dissociation
amplitude into a product of a purely two-body excitation amplitude
and a quasi-two-body scattering part of the center of mass of the
projectile has the consequence that all PDA effects are eliminated.

To overcome this problem we have proposed an approach which makes
use of a genuine three-charged particle wave function to describe
the final state. The latter is solution of the three-body \SE\!\!,
albeit only in an asymptotic region of configuration space which,
however, is relevant for breakup processes. Because of this our
estimation of three-body final-state interaction effects is reliable
only for small scattering angles. For larger angles a more accurate
three-body Coulomb scattering wave function has to be used which is
not yet available. Therefore, further investigations of the PDA
effects are called for. Nevertheless, we have shown that already the
use of even an only asymptotically correct wave function for three
charged particles in the continuum is beneficial for extracting
reliable values of astrophysical factors from Coulomb breakup
reactions, or at least can help pinpointing at kinematic conditions
in which PDA effects should play a negligible role. We finally
mention that the discovery of a low lying $E1$-strength in
neutron-rich nuclei and the demand to determine astrophysical
$S$-factors of radiative capture processes, as for
${}^7\textrm{Be}(p,\gamma){}^8\textrm{B}$, inspires new
calculations, possibly with new computational techniques, of Coulomb
breakup processes. Thus, Coulomb dissociation provides comprehensive
opportunities to study the interaction of exotic nuclei with
photons.

We have not mentioned the Coulomb breakup through excitation of a
resonance state of the projectile which is expected to lead to
interesting results in experimental as well as theoretical respect.

Among the problems still to be solved are (i) from the experimental
side the accurate determination of the relative energy of the
fragments, (ii) from the theoretical side the extrapolation of the
astrophysical factors to zero relative energy, the accurate
determination of the $E2$ contribution, relativistic corrections and
the nuclear interference terms in the cross section which become
more and more important as the scattering angle increases,
especially if it approaches the grazing angle.

\section*{Acknowledgments} This work was supported by the U.\ S.\ DOE under
Grant No.\ DE-FG03-93ER40773, by NSF Award No.\ PHY-0140343, by
Deutscher Akademischer Austauschdienst (DAAD) and by the Center of
Sciences and Technology of Uzbekistan, Project No $\Phi$2.1.44. B.\
F.\ I.\ would like to thank the Cyclotron Institute of Texas A\&M
University and the Institut f\"ur Physik der Universit\"at Mainz for
the hospitality extended to him during his visits.


\end{document}